# Experimental and numerical simulation study on the thermal performance of building envelope structures incorporating the solid-solid phase change material

Q. Ye[a*], L. Ba[b], G.T.M. Nguyen[c], R. Absi[b], B. A. Ledésert[a], G. Dosseh[c] and R.L. Hebert[a*]

[a]*Institut des Sciences de la Terre de Paris, CY Cergy Paris Université, 95000 Neuville-sur-Oise, France*
[b]*ECAM-EPMI, 13 Boulevard de l'Hautil, 95092 Cergy Pontoise, France*
[c]*LPPI, CY Cergy Paris Université, 5 Mail Gay Lussac, Neuville sur Oise F-95031, France*

## Abstract

This work is an experimental and numerical study of the thermal performance of building envelope structures incorporating a solid-solid phase change material (S-S PCM), consisting in a cross-linked polyurethane designated as PUX-1500-20. This S-S PCM is capable of storing and releasing thermal energy via phase transitions within the human comfort temperature range, facilitating the temporal and spatial transfer of solar energy for optimizing energy efficiency. The primary aim of this work is to integrate the S-S PCM into hollow bricks used in building envelopes and to evaluate their thermal inertia through both experimental testing and numerical simulation. The experimental results demonstrate that the integration of the PCM effectively delays and decreases the indoor temperature peak. The simulation results also show that the incorporation of the S-S PCM into hollow bricks gives rise to a phase shift of 7 hours and a decrement factor of 0.38. In comparison with the thermal behavior of the building envelopes (hollow brick) without PCMs, our results provide convincing evidence of the important thermal inertia of these structures incorporating the PCMs, revealing their significant potential in reducing energy consumption of building.

| **Nomenclature** | |
|---|---|
| ATFR | Average temperature fluctuation reduction |
| BPCM | Bio-based phase change material |
| CHBs | Concrete hollow blocks |
| CLSC | Clastic lightweight shale ceramsite |
| CM | COMSOL Multiphysics |
| CNT | Carbon nanotube |
| DF | Decrement factor |
| DSC | Differential scanning calorimetry |
| EGA | Expanded glass aggregate |
| EXP | Experimental study |
| FEM | Finite Element Method |
| HFR | Heat flux reduction |
| HMDI | Hexamethylene diisocyanate |
| HVAC | Heating, ventilation and air conditioning |
| LA-SA/$Al_2O_3$/C | Lauryl alcohol and Stearic acid/Aluminum Oxide |
| LCCCs | Lightweight cementitious cellular composites |
| MF | Melamine-formaldehyde |
| mPCM | Micro-encapsulated phase change material |
| [NCO] | Isocyanate group |
| NUM | Numerical study |
| [OH] | Hydroxyl group |
| PCM | Phase change material |
| PEG | Polyethylene glycol |
| PUX | Cross-linked polyurethane |
| PVC | Polyvinyl chloride |
| RMTR | Room maximum temperature reduction |
| S-L PCM | Solid-liquid phase change material |
| S-S PCM | Solid-solid phase change material |
| SSTF | Stearin of sheep tail-fat |
| TA | Thermal analysis |
| TL | Time lag |
| UF | Urea formaldehyde resin |
| WCF | Waste chicken feather |
| Cp | Apparent heat capacity (kJ/(kg·K)) |
| $C_p$ | Specific heat capacity (kJ/(kg·K)) |
| f | Decrement factor |
| $\Delta H_{fusion}$ | Latent heat of fusion (Enthapy of fusion) (kJ/kg) |
| $\Delta H_{crystallization}$ | Latent heat of crystallization (Enthapy of crystallisation) (kJ/kg) |
| k | Thermal conductivity (W/(m·K)) |
| Q | Volumetric heat source ($W/m^3$) |
| $\rho$ | Volumetric mass ($kg/ m^3$) |
| $T_{crystallization}$ | Crystallization temperature (°C) |
| $T_{fusion}$ | Melting temperature (°C) |
| $T_{input}$ | Input temperature (°C) |
| $T_{output}$ | Output temperature (°C) |
| $\nabla \cdot \vec{q}$ | Divergence of heat flux ($W/m^3$) |
| ∇T (dT/dx) | Temperature gradient (°C/m) |

# 1. Introduction

Energy and environmental challenges are among the most critical issues that humanity is facing today. With the industrial development, population boom, climate change and the rising demand for thermal comfort in buildings, the sector of construction and building has become the largest consumer of energy in recent decades, accounting for approximately 40% of the world's annual energy consumption [1][2]. Over 27% of this amount comes from the household energy consumption, primarily for heating, ventilation, and air conditioning (HVAC) systems [4][5]. Due to massive excessive energy demand, this sector is a major contributor to global carbon emission [6][7]. To reduce the high energy consumption of building operation and the associated carbon emissions, one of the main challenges for future buildings is the rational design of building envelopes. This involves insulation improvements as well as using renewable energy (e.g. solar energy). Developing passive systems for thermal energy storage can also contribute to reducing the energy demand for cooling and heating while maintaining a thermally comfortable indoor environment. A proposed solution is to integrate smart materials into building envelopes.

Phase change materials (PCMs) are considered to be one of the most promising smart materials due to their ability to exchange (absorbing and delivering) thermal energy with the surrounding environment through phase transitions [8][9][10]. This property enhances the thermal inertia of buildings thus optimizing the thermal comfort of building occupants, reduces indoor temperature fluctuations, improves energy efficiency which allows downsizing cooling and heating systems. These benefits contribute to lowering greenhouse gas emission and therefore mitigating global warming [11].

Although PCM technology has exhibited robust performance in both theoretical models and laboratory experiments, its translation to practical applications remains fraught with challenges. Achieving a homogeneous and stable incorporation of PCMs into building components, while preserving their functional integrity over extended periods and under diverse climatic conditions, presents a significant technical obstacle. Furthermore, the inherent variability of building materials and the complexity of construction methodologies require ongoing advancements in PCM encapsulation techniques and composite integration strategies to achieve optimal synergistic performance. Especially in the context where the majority of PCMs available on the market are solid-liquid phase change materials (S-L PCMs), which must address the critical issue of leakage prevention. Currently, the most popular integration methods are known as encapsulation and shape stabilization [12].

In the encapsulation method, a dense material forms a coating around the core PCM as capsule. This approach can be further subdivided into microencapsulation and macroencapsulation, with distinctions that extend beyond mere size differences. Microencapsulation entails the random and uniform distribution of numerous PCM-filled capsules within the matrix material, whereas macroencapsulation typically integrates PCM as an integral part of the structural system, for example, as a layer embedded within a wall or as a panel affixed to a surface [12]. Representative research summaries [13-17] from the past five years are presented in **Table 1**. The results demonstrate that the incorporation of microencapsulated PCMs can significantly enhance the thermal inertia of buildings and improve indoor thermal comfort. However, the encapsulation process tends to increase the crystallization temperature and reduce the latent heat capacity. In addition, microencapsulation involves higher technical complexity and increased production costs, which may limit its large-scale application in the building sector [12].

| Typical Studies on Microencapsulated PCMs as Building Materials | | | | | | |
|---|---|---|---|---|---|---|
| Ref. | Configuration | | | Studied parameters | Study type | Key findings |
| | PCM | Integration material | Integration method | | | |
| Alakara et al. [2025] [13] | n-octodecane (90% purity material) | Mortar mixtures | Cement, water, and mPCM solution, sand mixed in sequence to form a homogeneous mix, then cast into block molds and compacted. | Thermal performance, mechanical stability, and flame retardancy. | EXP | Cement blocks integrated with boric acid–formaldehyde shelled n-octadecane mPCMs exhibit excellent thermal, mechanical, and flame-retardant properties; silver coating on microcapsules further enhances thermal conductivity and flame retardancy. |
| Lajimi et al. [2023] [14] | Paraffin RT25 | Concrete | The micro-encapsulated phase change concrete (PCC) applied as a 3 cm thick coating layer on the exterior ceiling wall. | Mechanical stability, thermal performance, and flame retardancy. | NUM | Walls with PCC show improved flame retardancy, mechanical strength, and thermal comfort, reducing AC cooling loads by 20% compared to traditional glass wool insulation. |
| Kirilovs et al. [2020] [15] | Bio-PCM-S50 (Commercial PCM developed by MikroCaps Ltd, Slovenia ) | Industrial hemp (Cannabis sativa L.) and Urea formaldehyde resin (UF) | Commercial urea-formaldehyde resin adhesive was used as adhesive to directly mix PCMs with hemp for the production of thermal insulation wallboards | Thermal conductivity and latent heat | EXP | Thermal conductivity ranged 0.064–0.074 W/(m·K), within commercial hemp fiber insulation standards; 5% mPCM increased specific heat by 62% to 2.369 J/(g·K). |
| Xue et al. [2024] [16] | Nextek 18D, a paraffin-based commercial product. | Plaster mortar | The mPCM-plaster composite layer was applied to the gypsum board through a spreading process. | Thermal performance at varying mPCM concentrations | EXP and NUM | PCM concentration has limited impact; optimized distribution improves thermal efficiency. |
| Wu et al. [2021] [17] | Paraffin wax | Cementitious mortar and concrete | mPCM with a melamine-formaldehyde (MF) shell were directly incorporated into cementitious mortar to: 1) cast bulk specimens ;2) fabricate lightweight cementitious cellular composites (LCCCs) using 3D printing-assisted techniques. | Mechanical and thermal insulation properties | EXP and NUM | Excellent thermal insulation performance with no significant reduction in compressive strength compared to conventional cement-based materials |

***Table 1:*** *Summary of studies on microencapsulated PCMs in building materials (EXP: experimental study; NUM: numerical study).*

Compared to microencapsulation, macroencapsulation provides a broader operational scope and a wider range of material selection options. Typical studies [18-24] conducted over the past five years are listed in **Table 2**. Macroencapsulation is more dependent on specific structural designs, which poses challenges for large-scale commercial processing and often requires more complex on-site installation. Nevertheless, owing to the continuous porosity of the matrix material, macroencapsulation is capable of achieving a significantly higher PCM content, thereby offering enhanced energy storage density. Moreover, since macroencapsulation units are generally used as independent structural elements, they typically do not compromise the mechanical performance of the matrix material or the overall structural integrity of the building [12].

| Typical Studies on Macroencapsulated PCMs as Building Materials | | | | | | |
|---|---|---|---|---|---|---|
| Ref. | Configuration | | | Studied parameters | Study type | Key findings |
| | PCM | Integration material | Integration method | | | |
| Rathore et al. [2020] [18] | SavE® OM37 (commercial inorganic chemical-based PCM developed by Pluss Advanced Technologies Pvt. Ltd.) | Aluminum Alloy 8011 tube (Ø16.7 mm × 900 mm, 0.5 mm thick) | PCM powder directly filled into aluminum alloy tubes and embedded in wall and roof structures. | Thermal inertia and cooling load | EXP | Integration of macro-encapsulated PCM reduced peak temperatures by 7.19–9.18%, thermal amplitude by 40.67–59.79%, and peak heat flux by up to 41.31%, with a 38.76% drop in cooling load and 60–120 min delay in peak temperature. |
| Henieg-al et al. [2021] [19] | paraffin wax (Ameeriah Petroleum Refinery Company, Alexandria, Egypt) | Concrete hollow blocks (CHBs) | CHBs cavities filled with PCM blends using pumice or mortar at different proportions. | Thermal inertia and electrical consumption energy saving | EXP | PCM-pumice and PCM composite mortar significantly enhance thermal energy storage and comfort, reducing building cooling loads by over 25%. |
| Abdulm-unem et al. [2022] [20] | Paraffin wax type Heptacosane ($CH_3(CH_2)_{25}CH_3$) | Hollow PVC plastic panel | Finely ground WCF was mixed with molten PCM then poured into PVC panel cavities at 50 °C to form integrated insulation panels, which were installed on the interior surfaces of walls and ceilings. | Thermal, mechanical and acoustic performance, electricity cost saving | EXP | PVC panels filled with PCM and WCF used as inner wall materials exhibit strong mechanical properties and reduce cooling load, noise, and electricity costs by 9.6%, 9%, and 22.5%, respectively. |
| Abdulm-unem et al. [2025] [21] | Stearin of sheep tail-fat (SSTF), a bio-based PCM (BPCM). | Transparent polycarbonate container (12 × 12 × 2 cm, square shape) simulating a building envelope. | SSTF directly melted in the container (120°C); CNT added gradually and dispersed by stirring. | Melting behavior | EXP and NUM | SSTF with CNT shows ~13.7% increase in heat transfer efficiency and ~7.5% increase in thermal storage rate; effectively reduces building heating/cooling loads, lowers energy consumption, and decreases greenhouse gas emissions for sustainable thermal comfort control. |
| Khan et al. [2020] [22] | Paraffin wax type P56-58 supplied by the MERCK | Cavity constructed from brick and plaster | PCM separately filled in two cavities (brick and plaster) located inside the water bath box. | Impact of PCM on heat transfer | EXP | PCM application reduces indoor temperature fluctuations, enhancing thermal comfort; higher efficiency when placed near heat source. |
| Al-Yasiri et al. [2021] [23] | Paraffin wax | Hollow galvanized steel panel (total thickness: 10 mm; wall thickness: 0.4 mm) | Panels filled with melted PCM and integrated at different roof positions | RMTR, ATFR, DF, TL and HFR | EXP | Indoor temperature significantly reduced and fluctuations smoothed for all PCM positions; highest efficiency achieved near outer roof layer. High-melting PCM effectively utilized at elevated external temperatures. |
| Arici et al. [2020] [24] | Paraffin RT-category [25] | External building wall | Integration of PCM with varying thicknesses at different wall locations: between insulation and exterior plaster, and between interior plaster and concrete. | Energy savings, DF, and TL under different PCM locations, $T_{fusion}$, thicknesses. | NUM | PCM placement near exterior favors heating savings; near interior favors cooling savings; Optimizing PCM thickness and melting temperature is key to cost-effective passive building energy systems. |

***Table 2:*** *Summary of studies on macroencapsulated PCMs in building materials (EXP: experimental study; NUM: numerical study).*

Shape stabilization is another widely used incorporation technique of PCMs. This method involves impregnating PCMs into a porous supporting material to form a shape-stable PCM composite. The most commonly used processing methods are vacuum impregnation and melt (direct) impregnation, along with other approaches such as self-acting combinations. **Table 3** summarizes the key research findings on this topic published over the past five years [26-30]. The results indicate that the factors such as type of raw materials, particle size, material ratio, additives etc. have a significant impact on the performance of PCM. Especially, the significant reduction of mechanical strength and phase transition enthalpy remains a critical issue, which is primarily influenced by the PCM mass fraction and phase transition rate. Minimizing the negative impact on the merger process is still a research focus [12].

| Typical Studies on Shape Stabilized PCMs as Building Materials | | | | | | |
|---|---|---|---|---|---|---|
| Ref. | Configuration | | | Studied parameters | Study type | Key findings |
| | PCM | Integration material | Integration method | | | |
| Yang et al. [2020] [26] | Binary PCM based on Ceramsite: LA-SA/$Al_2O_3$ | Concrete | PCM Directly Incorporated into concrete to fabricate composite PCM blocks | Optimal ratio, stability, thermal and mechanical properties | EXP | Optimal ratio: LA (82 wt%), SA (18% wt%), and $Al_2O_3$ ( 0.5% wt%); good chemical and thermal stability; good thermal inertia. |
| Qu et al. [2020] [27] | Paraffin/fumed silica composite | Cement mortar and foaming agent | Raw materials were mixed sequentially and cast into foam concrete blocks | Optimal ratio, stability, thermal and mechanical properties | EXP | Composite PCM with 45% paraffin showed the most stable structure, with significantly enhanced thermal conductivity and heat storage capacity |
| Yousefi et al. [2021] [28] | Pure Temp 23, a bio-based commercial PCM from Pure Temps LLC. | Cement mortar | Vacuum impregnation of PCM into EGA pores, followed by fly ash coating, then mixed with cement mortar. | Structure, thermal and mechanical properties | EXP | Uniform structure; significantly improves thermal performance but reduces mechanical strength. |
| Shen et al. [2021] [29] | Semi-refined paraffin from China National Petroleum Corporation. | Concrete | Paraffin absorbed into clastic lightweight shale ceramsite (CLSC) to prepare PCM-CLSC aggregates, then integrated into concrete to form blocks. | Compatibility, absorption, thermal, and mechanical performance | EXP | Good compatibility; 25.8% absorption rate; significantly enhanced thermal storage, but greatly reduced compressive strength. |
| Liao et al. [2024] [30] | Paraffin from Zhongmiao Petrochemical Co., Ltd. | Cement mortar | Liquid paraffin impregnated into coal gasification slag powder to prepare shape-stabilized PCM, then mixed with cement mortar and cast into blocks. | Physical, mechanical, and thermal properties | EXP | Composite with 30 wt% paraffin reduces cooling load; thermal conductivity, compressive strength, flexural strength, fluidity, and density decrease by 46.7%, 66.2%, 16.8%, 5.3%, and 20.4%, respectively. |

***Table 3:*** *Summary of studies on shape stabilized PCMs in building materials (EXP: experimental study; NUM: numerical study).*

In 2017, Harlé et al. developed and patented a solid-solid phase change material (S-S PCM) suitable for building application, named PUX-1500-20, which is a cross-linked polyurethane [31][32][33]. This material presents significant advantages: i) phase change temperature within the human comfort range ($T_{fusion}$ =38°C, $T_{crystallization}$ = 22.3°C); ii) high latent heat ($\Delta H_{fusion}$ = 91J/g, $\Delta H_{crystallization}$ = 89J/g); iii) no risk of leakage and minimal volume expansion rate due to the solid-solid phase change; iv) non-toxic and non-corrosive; v) high hardness (shore D: 30 ± 1); and vi) long term durability (thermal and chemical stability).

In their subsequent studies [34][35][36], PUX-1500-20 was directly incorporated into gypsum, cement paste, and mortar suspensions as a solid adjuction. Although these studies demonstrated improved thermal performance, they also revealed a significant reduction in mechanical strength, primarily due to the partial water solubility of the PCM in hydrous binder systems.

These studies predominantly relied on binder-based integration methods, where the PCM was inevitably exposed to moisture during mixing, leading to partial dissolution and compromised mechanical performance of the resulting composites. To date, there is still a lack of systematic research on integration strategies that can simultaneously preserve both the thermal and mechanical properties of the composite material in practical applications.

Building upon this prior work, the present study draws inspiration from existing approaches where solid-liquid paraffin-based PCMs were embedded into structural cavities of building elements [19][22]. A novel integration strategy is proposed here: the direct integration of S-S PCM into structural elements such as hollow bricks, in the form of “macroencapsulation”. This method is not only straightforward to pouring but also structurally advantageous. First, the brick shell naturally prevents direct contact between the internal PCM and moisture. Second, the placement of PCM within the cavities, without mixing it into the brick matrix, ensures that the composition and mechanical properties of the brick remain unaffected. Consequently, this approach maintains the mechanical stability of the host material while delivering effective thermal performance.

Furthermore, by combining experimental testing with numerical simulation, this study offers a comprehensive evaluation of the thermal behavior of the integrated system, laying a theoretical and empirical foundation for future optimization strategies.

## 2. Experimental section

### *2.1 Synthesis and integration of the PCMs with hollow bricks*

The raw materials polyethylene glycol (PEG)-1500 and glycerol (purity ≥ 99%) were purchased from Carl Roth GmbH + Co, and hexamethylene diisocyanate (HMDI) MERCK MILLIPORE from VWR (purity ≥ 99%). All reagents were used as received.

The synthesis was realized in a 10L three-necked glass reactor under an argon atmosphere. A total of 3kg of PEG-1500 and 24.56g of glycerol (corresponding to 20% of glycerol hydroxyl function, compared with that of PEG) were added first into the reactor placed in a 60°C water bath. The mixture was stirred until the PEG was completely melted and the reactants were homogenized before the incorporation of HMDI to realize a one-step solvent-free synthesis. The molar ratio of isocyanate group/hydroxyl group ([NCO] ⁄ [OH]) was fixed at 1. After incorporating all the reactants, the mixture was stirred at 60 °C for 45min then poured directly into hollow bricks used in building envelopes. The polymer-filled brick matrix subsequently underwent a post-curing process in an oven at 100 °C for 5 hours. Further details regarding the synthesis of PUX-1500-20 are available in the study by Harlé et al. (2020)[32].

### *2.2 Characterization*

#### *2.2.1Differential scanning calorimetry (DSC)*

The DSC analysis was performed in a differential scanning calorimeter, TA instruments Q100, in order to determine the transition temperatures, latent heat and specific heat capacity of the PCM.

The DSC analysis was performed on PCM powdered samples of 2 ~ 5mg placed into aluminum hermetic crucibles as well as aluminum hermetic crucible with vacuum for reference. A first heating-cooling cycle from -10°C to 80 °C was realized to erase the thermal heritage of the sample. Samples were then analyzed from -10 °C to 80 °C, with cooling and heating rates of 5 °C/min. Each sample was analyzed three times, and the results presented in **Table 4** correspond to the mean values of the three measurements.

#### *2.2.2Measurement of density*

The determination of sample density by triple weighing method was carried out with its saturated aqueous solution instead of water only because of its partial solubility in pure water. The same sample is measured five times to minimize error.

The result presented in **Table 4** is the mean value. **Table 4** lists all the parameters necessary for the numerical simulation.

| Property | Value |
|---|---|
| Density (kg/m$^3$) | 1140 ± 0.005 |
| Thermal conductivity (W/(m·K) | |
| Semi-crystalline phase | 0.231 [32] |
| Amorphous phase | 0.231[39] |
| Phase transition temperature (°C) | |
| Melting point | 38 ± 1.3 [32][39] |
| Freezing point | 22.3 ± 0.4 [32] |
| Specific heat capacity (J/(g·K)) | |
| Semi-crystalline phase | 1.31 ± 0.003 |
| Amorphous phase | 1.51 ± 0.002 |
| Latent heat (J/g) | |
| Fusion | 91 ± 2 [32] |
| Crystallization | 89 ± 2 [32] |

***Table 4:*** *Thermo-properties of the PCM PUX-1500-20.*

### *2.3 Experimental test*

#### *2.3.1Hotplate*

The experimental test was performed on a high-precision hotplate PZ28-1 from Harry Gestigkeit®, with an accuracy of ±0.1°C and dimensions of 20*28 cm. The temperature data was collected using an Agilent 34970A data acquisition switch unit equipped with K-type thermocouples. A scan step size of 2 minutes was used.

#### *2.3.2Brick*

The hollow bricks used in this study were purchased from a local building material store (Leroy Merlin). They are commercially available standard terracotta bricks with dimensions of 38.5*25.3*5cm.

The bricks were resized while maintaining the original length-to-width ratio for practical reasons: (i) to reduce the amount of PCM required for synthesis, and (ii) to ensure compatibility with the hotplate dimensions used for experimental testing. The dimensions of the resized brick are 23.6*15.5*5cm.

Subsequently, five thermocouples were emplaced into the brick as well on their surface to collect temperature evolution data at six different locations (see **Figure 1**). Thermocouples 1 and 5 were directly glued to the surface of the bricks using strong adhesive. For the PCM-filled bricks, thermocouples 2, 3, and 4 were directly implanted into holes of corresponding depth, which were pre-drilled using an electric drill, and fixed using softened PCM. For the bricks without PCM, thermocouples 2, 3, and 4 were implanted in the corresponding positions after being fixed with small pieces of foam material.

In order to simulate a real wall environment, a 5mm thick cement layer is coated on both ends of the brick matrix before testing. **Figure 2** presents photographs of the original brick, the resized brick, and the brick with emplaced thermocouples and a surrounding cement coating, respectively.

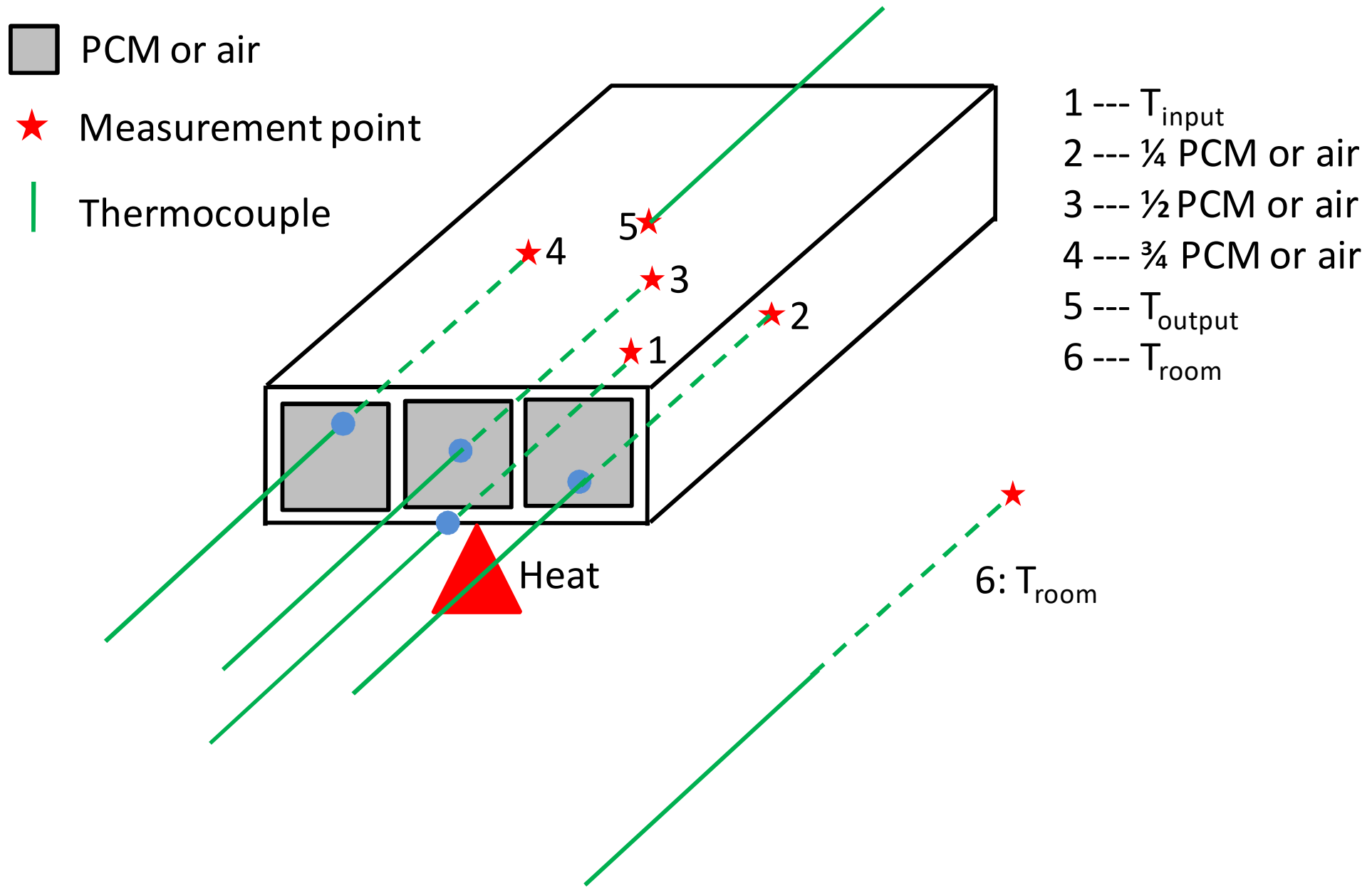

***Figure 1:*** *Illustration of the experimental specimen and temperature data collection points.*

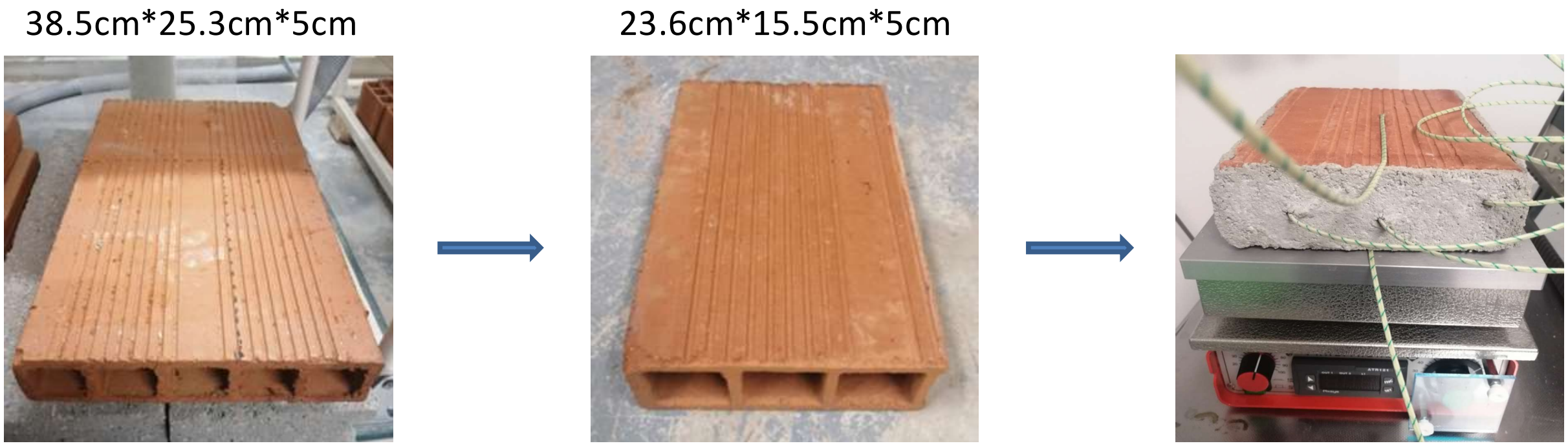

***Figure 2:*** *Photographs of the original brick (left), the resized brick (middle), and the brick with emplaced thermocouples and a surrounding cement coating (right).*

#### *2.3.3Experimental methods*

The experimental test was conducted under the following two heat source conditions:

1) A constant temperature of 50°C, corresponding to the peak temperature typically recorded in recent summers on the air-exposed side of residential building exterior walls in France [37], was applied to one side of the brick, with/without PCM, for duration of 6h30, followed by natural cooling at room temperature.

2) A constant temperature of 80°C, significantly exceeding the phase change temperature, was applied to one side of the brick, with/without PCM, for 24h to facilitate a clearer observation of the phase change process.

### *2.4 Numerical simulation*

The numerical simulation of heat transfer in this study is performed using COMSOL Multiphysics - 6.2 (CM) [38], which uses the Finite Element Method (FEM) to solve the heat transfer equation. Specifically, in the FEM framework, the computational domain is discretized into a finite number of subdomains, referred to as elements, through a meshing process. Within each element, the field variables, such as temperature, are approximated using polynomial basis functions, known as shape functions. The governing equations are then reformulated into their weak form via the weighted residual method, typically employing the Galerkin approach. This process results in a system of algebraic equations, which are subsequently solved numerically to obtain an approximate solution for the entire computational domain. To accurately simulate phase change phenomena in heat transfer analyses, the Heat Transfer Module provides two approaches. The Phase Change Material feature implements the apparent heat capacity formulation, accounting for enthalpy of phase change and variations in material properties. This method also enables the modeling of volume and topology changes. Alternatively, the Phase Change Interface feature models phase change based on the Stefan energy balance condition to compute the velocity of the interface between two phases that may have different densities. Combined with deformed geometry, this approach is very efficient and effective in cases without topology change [38].

In this work, the computational model accounts for conduction-dominated heat transfer, latent heat effects associated with phase transitions, and interfacial thermal resistance between the PCM and brick matrix. The heat transfer is considered to be one-dimensional in the direction of thickness of the bricks (from outdoor to indoor). The simulation results are influenced by the thicknesses of both the bricks and the PCM, whereas the height and width of the bricks have no impact on the outcomes. The constructed geometric model, shown in **Figure 3**, is designed based on the dimensions used in the experimental test. The thermo-physical properties of the PCM PUX-1500-20 [32] and the surrounding heat transfer medias are shown in **Table 4** and **Table 5**, respectively.

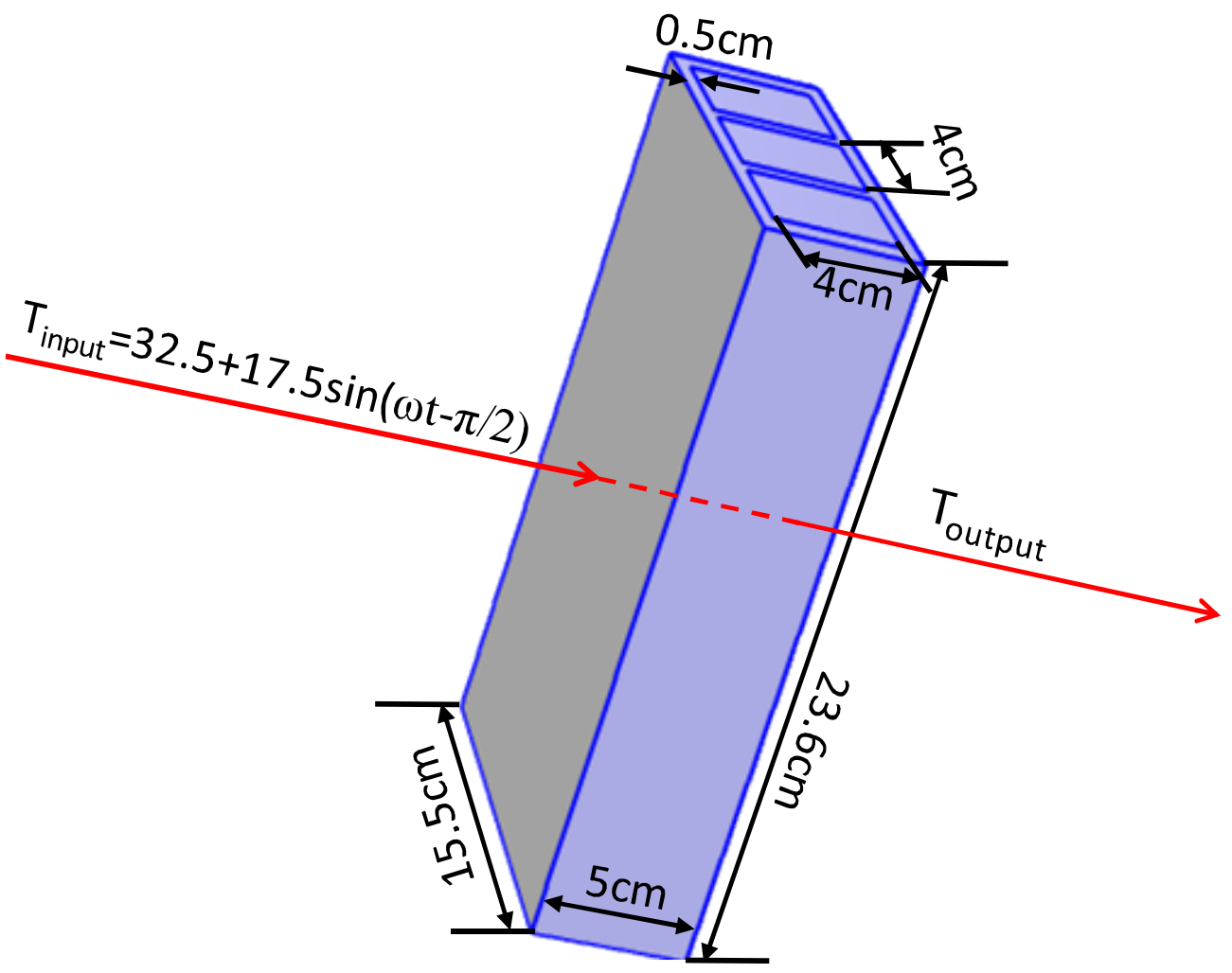

***Figure 3:*** *Schematic representation of the hollow brick which can be filled with PCM, the blue color indicates that all surfaces of the brick, except for the heating surface (grey), were insulated.*

| Material | Thermal conductivity (W/(m·K)) | Density ($kg/m^3$) | Specific heat capacity (kJ/(kg·K)) |
|---|---|---|---|
| Brick | 1.15 | 2300 | 0.92 |
| Air | 0.03 | 1.20 | 1.00 |

***Table 5:*** *Thermo-properties of the brick and air* [39].

To study this system dominated by heat conduction, the following assumptions are made to simplify the governing equation:

- The PCM is thermally homogeneous and isotropic.
- The PCM is initially in a solid semi-crystalline phase.
- Heat transfer through the materials is one-dimensional and occurs in an unsteady regime.
- Only conduction and phase change heat transfer are considered, while convection, radiation, and other heat transfer mechanisms are neglected.
- The thermo-physical properties of the PCM and other materials are assumed to be independent of temperature.

Consequently, the governing time-dependent energy conservation equation is simplified into the following heat conduction equation using the apparent heat capacity ($C_p$) method.

$$\rho\, C_p \frac{\partial T}{\partial t} + \nabla \cdot \vec{q} = Q$$

where $\rho$ is the volumetric mass ($kg/m^3$), $C_p$is the specific heat capacity (kJ/(kg·K)), $Q$ is the volumetric heat source ($W/m^3$), and $\nabla \cdot \vec{q}$ is the divergence of heat flux representing heat diffusion (conduction, $W/m^3$). According to Fourier's first law,

$$q = -k\nabla T$$

where k is the thermal conductivity (W/(m·K)), and $\nabla T$ is the temperature gradient: dT/dx (°C/m).

The imposed temperature is modeled using a sinusoidal approximation to simulate the actual 24-hour temperature fluctuations (ranging from 15°C to 50°C) experienced on the air-exposed side of residential building exterior walls during the hotter season in France in recent years [37]. Therefore, the imposed temperature equation is a sinusoidal function with a baseline offset of 32.5°C, amplitude of 17.5°C, angular frequency of ω (ω = 2πf = 2π/T, T =24h), and a phase shift of π/2, mathematically expressed as:

$$T_{input}(t) = 32.5 + 17.5 \sin(\omega t - \pi/2)$$

with the temperature T in degree Celsius (°C) and the time t in hours (h).

## 3. Results and discussion

### *3.1 Experimental test*

As mentioned in the experimental section, the experimental test was conducted under the following two heat source conditions: 1) a constant temperature of 50°C was applied to one side of the brick with/without the PCM for 6h30min, followed by natural cooling at room temperature; 2) a constant temperature of 80°C was applied to one side of the bricks, with/without the PCM, for 24h.

In the first case, the temperature evolution curves at each measurement point are shown in **Figure 4**. We can observe that in both the PCM-filled system and the system without PCM, the temperature decreases gradually from the heated surface outward, forming a thermal gradient. The closer to the heat source, the smaller the temperature difference between over equal distances. Interestingly, at the 1/4 and 1/2 positions, the temperatures in the PCM-filled brick are 0.9°C and 0.3°C higher, respectively, than compared to those at the same positions in the brick without PCM. In contrast, at the 3/4 position and on the surface opposite to the heat source, the temperatures are 1.7°C and 1.0°C lower, respectively. This indicates that the PCM absorbs and stores a substantial amount of latent heat during the phase transition process, thereby delaying the temperature rise on the side farther from the heat source. When the system reaches thermal equilibrium, the thermal buffering effect of the PCM leads to a significantly reduced temperature gradient within the brick. This thermal regulation behavior is primarily attributed to the high latent heat capacity of the PCM, which enables it to absorb and store considerable heat with minimal temperature fluctuation during the phase change. In practical applications, this suggests that the PCM can effectively retain a large amount of thermal energy within the internal structure of the brick, thereby reducing the heat transferred to the indoor environment. Consequently, PCM-enhanced bricks function as thermal energy storage units, helping to mitigate indoor temperature fluctuations, lower peak cooling demands, and improve overall energy efficiency during hot summer conditions.

On the other hand, when comparing the temperature variations during the heating and cooling phases between the two systems, it is evident that the rate of temperature change in the PCM-filled system is significantly lower than that of the non-PCM system in both phases. The PCM-filled system and the non-PCM system reach thermal equilibrium with the surrounding environment after heating in 5 hours and 1 hour, respectively, while cooling back to room temperature takes 8 hours and 6 hours, respectively. This demonstrates that the PCM-filled system possesses excellent thermal inertia, effectively reducing indoor temperature fluctuations.

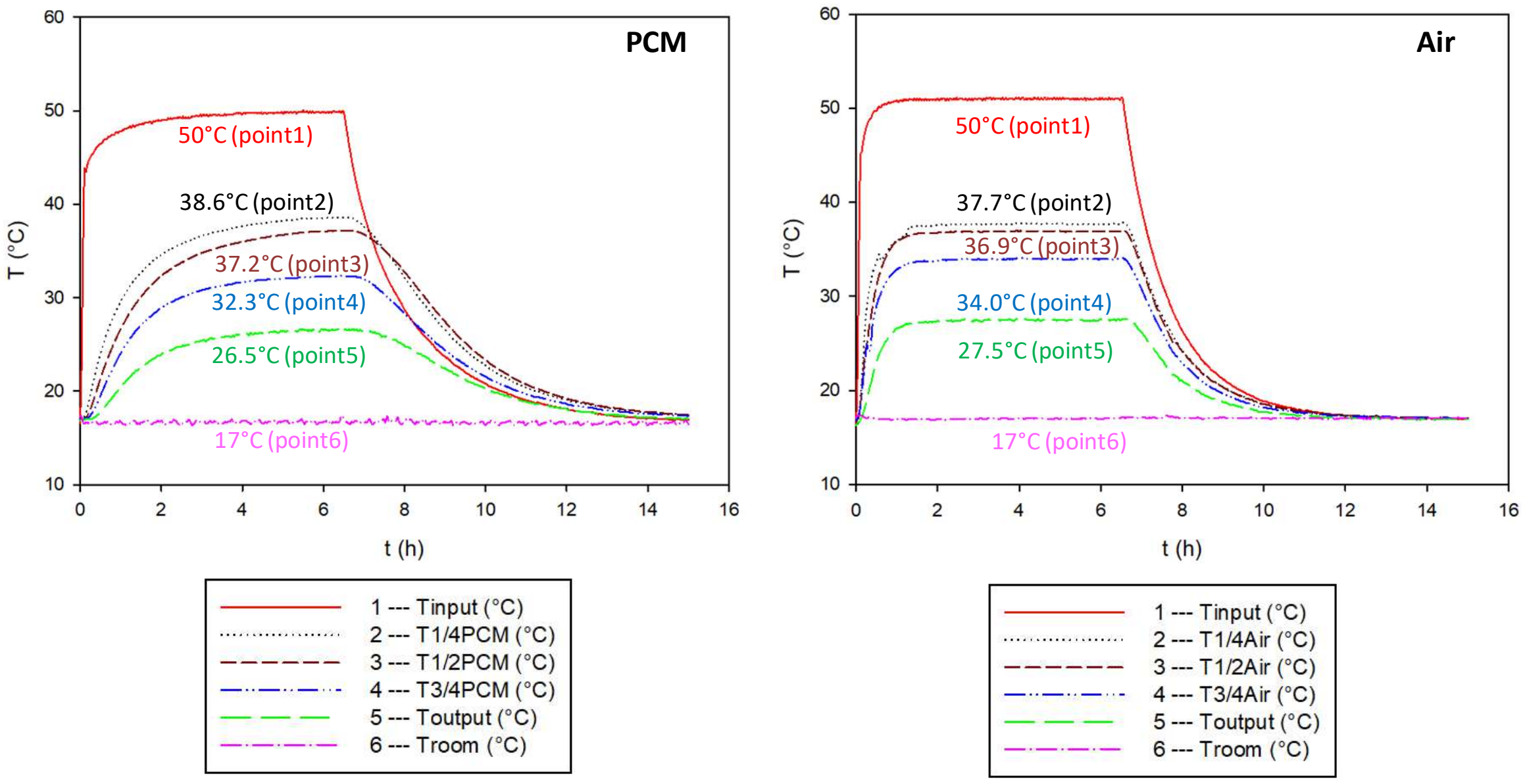

***Figure 4:*** *Temperature evolution over time for the PCM-filled brick (left) and the brick without PCM (right) during heating at 50°C for 6h30min, followed by natural cooling to room temperature.*

Similarly, when subjected to continuous heating on one side of the bricks with an 80°C heat source for 24 hours, the temperature at various measurement points in the PCM-filled brick rises more slowly compared to the corresponding points in the brick without PCM, due to the phase change effect of the material, as shown in **Figure 5**. A distinct phase change temperature plateau is observed (highlighted in blue on the left side of **Figure 5**), indicating the latent heat absorption during the phase transition process. It is evident that the farther a region is from the heat source, the slower the phase change occurs and the longer it takes to complete. It is observed also that in both the PCM-filled system and the system without PCM, the temperature decreases gradually from the heated surface outward, forming a thermal gradient. The closer to the heat source, the smaller the temperature difference between equal distances. At the 1/4 and 1/2 positions, the temperatures in the PCM-filled brick are 5.3°C and 5.1°C higher, respectively, compared to those at the same positions in the brick without PCM. In contrast, at the 3/4 position and on the surface opposite the heat source, the temperatures are 2°C and 1.5°C lower, respectively.

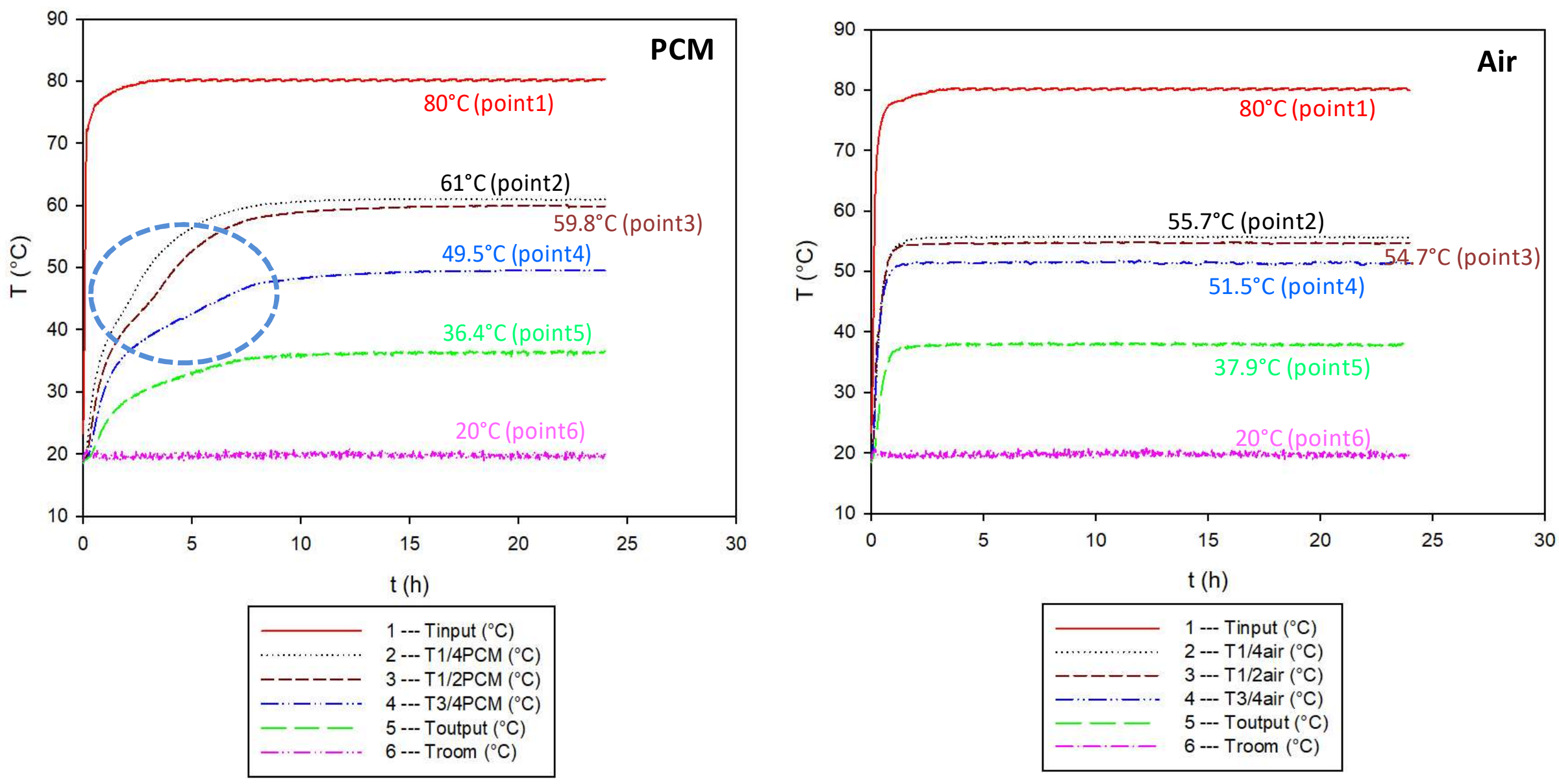

***Figure 5:*** *Temperature evolution over time for the PCM-filled brick (left) and the brick without PCM (right) during heating at 80°C for 24h, the blue circle indicates the temperature plateau caused by the phase change of the material.*

### *3.2 Numerical simulation*

In order to obtain results with relatively high resolution, a “fine” mesh was selected for the 3D geometry in this study, with element sizes ranging from a few tenths of a millimeter to approximately 1 millimeter, depending on the simulation domain. Meanwhile, a time step of 1h was applied. **Figure 6** presents the input and output temperature evolution curves for the PCM-filled brick and the brick without PCM, respectively.

During the thermal ramp-up phase, a noticeable reduction in the rate of temperature increase was observed in the PCM-filled system. This occurred between the 10th and 19th hours, corresponding to a temperature range of 33.36 °C to 37.88 °C. This reduced heating rate can be attributed to the latent heat absorption associated with the melting phase transition of the PCM. During this period, the supplied thermal energy was primarily consumed by the endothermic phase change process, rather than contributing to a rise in sensible temperature. As a result, the temperature of the system remained relatively stable despite continued heat input. The peak temperature in the PCM-filled system was 37.88 °C, indicating a delayed and dampened thermal response compared to the control system without PCM. In contrast, the non-PCM system exhibited a continuous and linear increase in temperature throughout the same period, reaching a peak of 38.75 °C at the 16.5th hour. The absence of a phase change mechanism in the control system meant that all incoming thermal energy directly contributed to sensible heating, leading to a faster and higher temperature rise. These observations highlight the significant thermal regulation capacity of PCM through latent heat storage, which plays a critical role in moderating temperature variations and enhancing thermal inertia under dynamic thermal loading conditions.

On the contrary, during the cooling process, the PCM brick exhibited a faster temperature decline than the non-PCM brick attributed to its higher thermal conductivity (0.231 W/(m·K) compared to 0.03 W/(m·K), which facilitates a more efficient reduction of indoor temperature under elevated thermal conditions. The PCM brick reached a minimum temperature of 24°C at the 28.5th hour, while the non-PCM brick reached 25°C at the 28.5th hour.

Consequently, the phase shift of the PCM-filled system is 2.5 hours higher than that of the non-PCM system. They are 7 hours and 4.5 hours, respectively. Moreover, it is observed that the amplitude was affected for the PCM-filled system. This result was determined by the decrement factor, which is the ratio between the two values of the amplitudes to that of the input temperature and that of the output temperature, as given by the following formula:

$$f = (T_{output\text{-}max} - T_{output\text{-}min}) / (T_{iutput\text{-}max} - T_{iutput\text{-}min})$$

The lower this value, the lower the thermal amplitude compared to the applied temperature, indicating that the heat transfer structure is more effective in insulating and stabilizing the indoor temperature. The calculated decrement factor is then 0.38 for the PCM-filled brick and 0.39 for the non-PCM brick. This reveals that PCM-filled brick has better thermal performance.

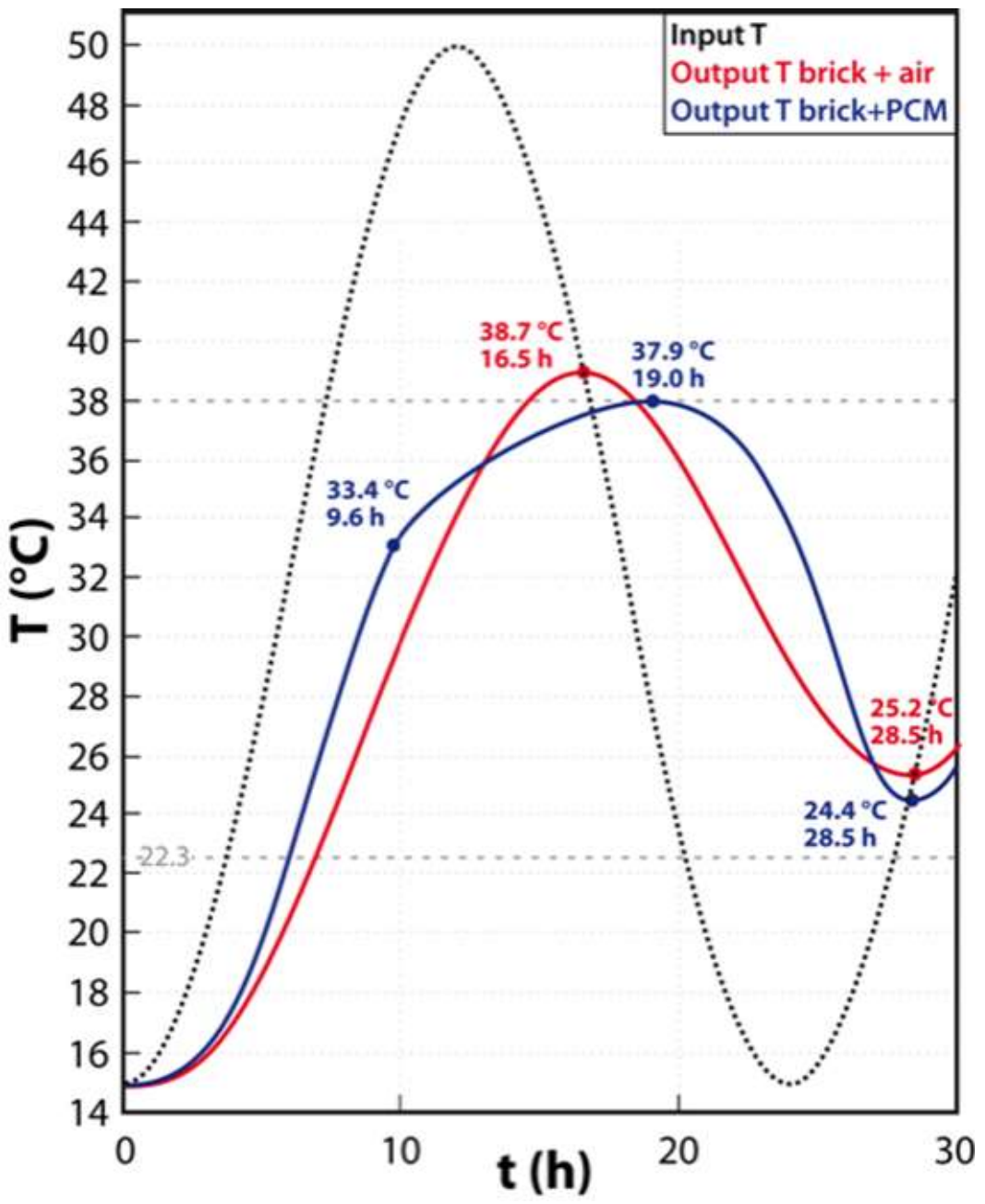

***Figure 6:*** *Input temperature curve (black dashed line) and output temperature curve for the PCM-filled brick (blue solid line) and the brick without PCM (red solid line).*

The detailed internal temperature evolution of the two types of bricks with respect to depth and time is presented as arc-shaped curves in **Figure 7** (PCM-filled brick) and **8** (non-PCM brick). From the right-side images of the two figures, it can be observed that the output temperature in both types of bricks shows little change during the first hour. From the second hour, the output temperature begins to rise rapidly. However, in the PCM-filled brick, between the 10th and 24th hours, the output temperature changes very slowly, as evidenced by the very narrow spacing between adjacent temperature curves. This is due to the phase change effect of the material embedded within the brick.

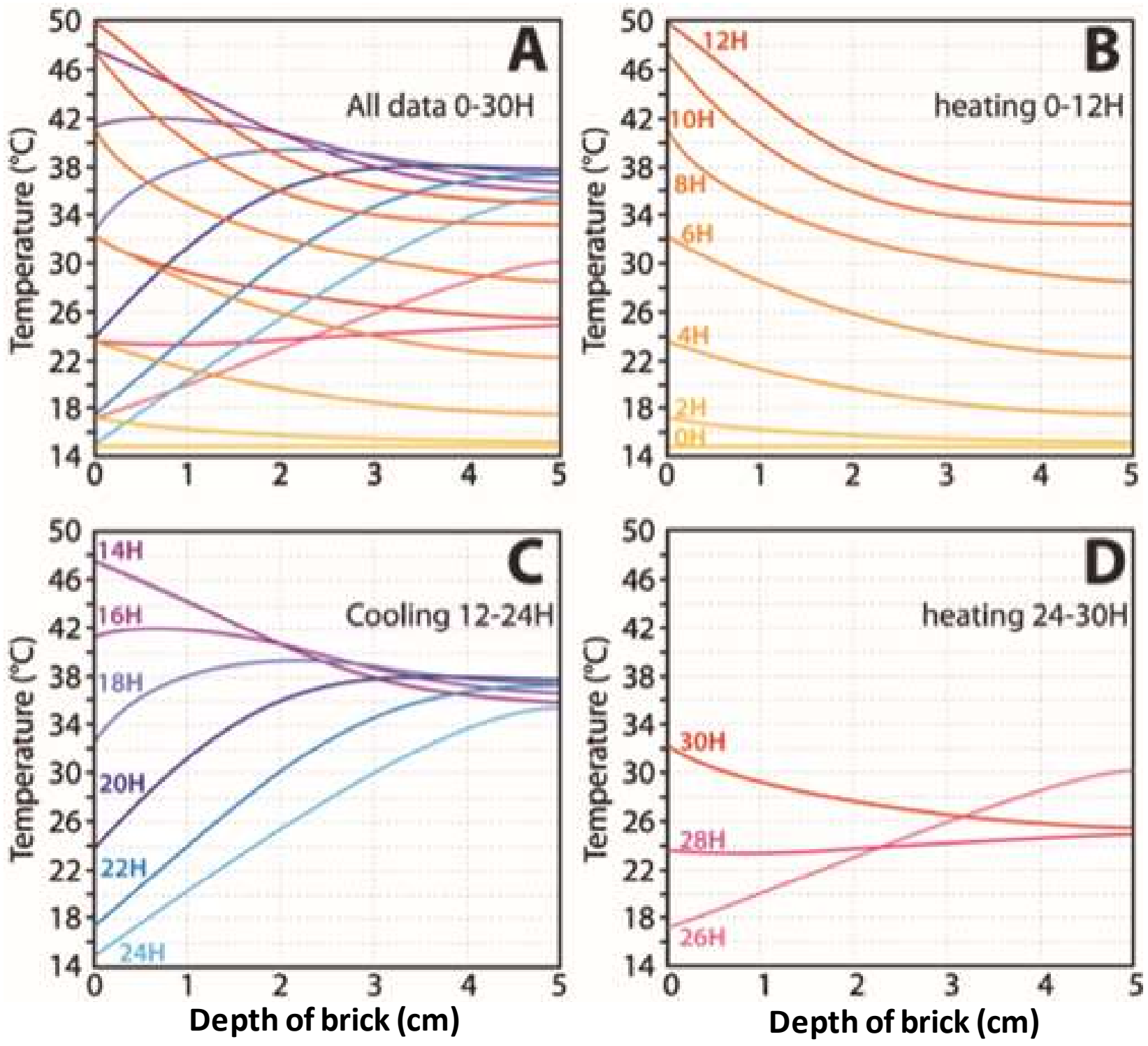

***Figure 7:*** *Temperature evolution with the depth of the PCM-filled brick every two hours.*

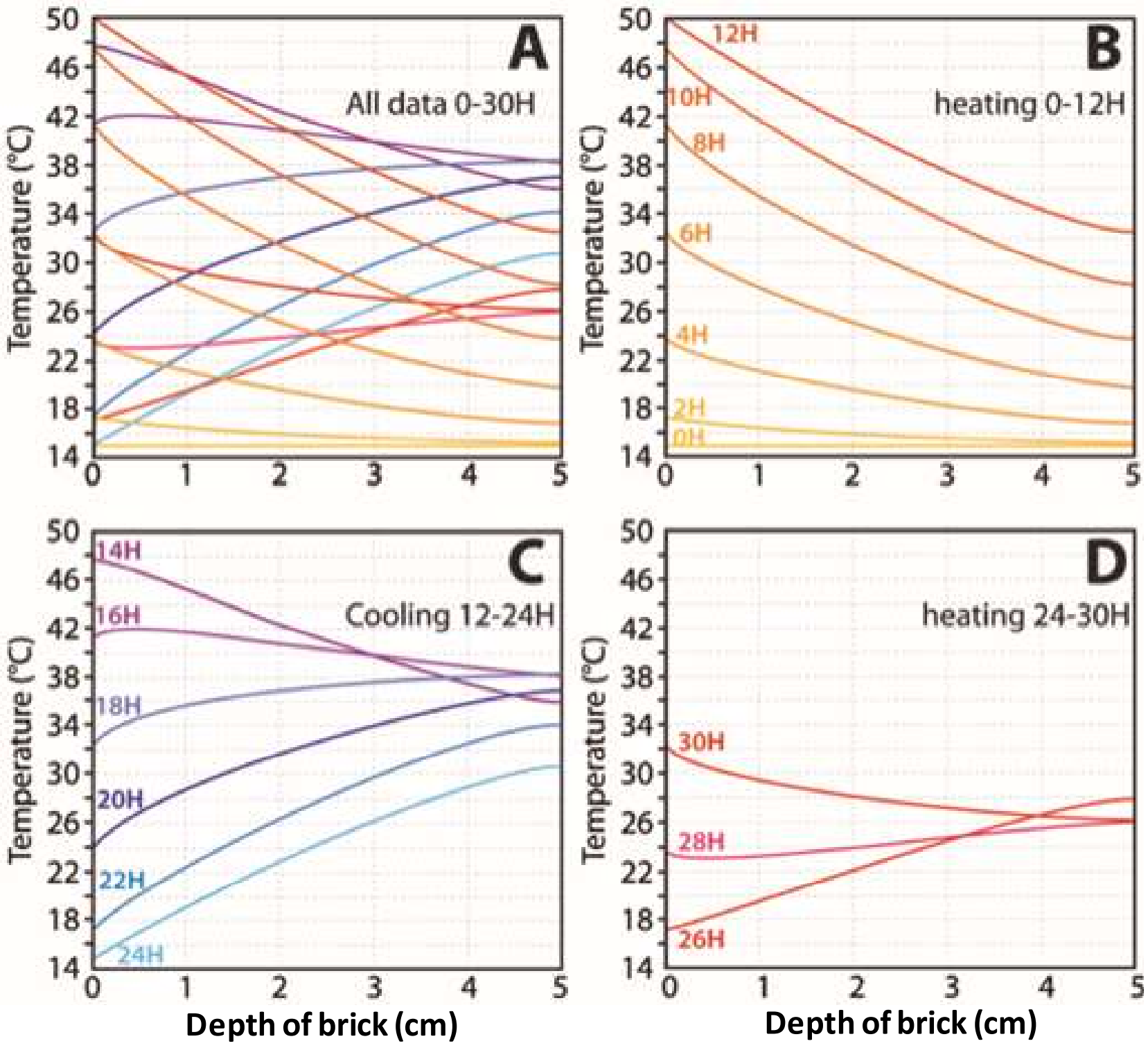

***Figure 8:*** *Temperature variation across the depth of the brick without PCM every two hours.*

To further visualize the comparison of the internal temperature evolution between the two types of bricks, the temperature evolution within the bricks is presented as slice section plots in **Figure 9** and **10**. In all diagrams where a heat source is applied, it can be observed that the temperature difference between the heated surface and the output surface is smaller in the PCM-filled system compared to the non-PCM system. This indicates that the PCM-filled system absorbs a significant amount of heat, resulting in a more stable output temperature.

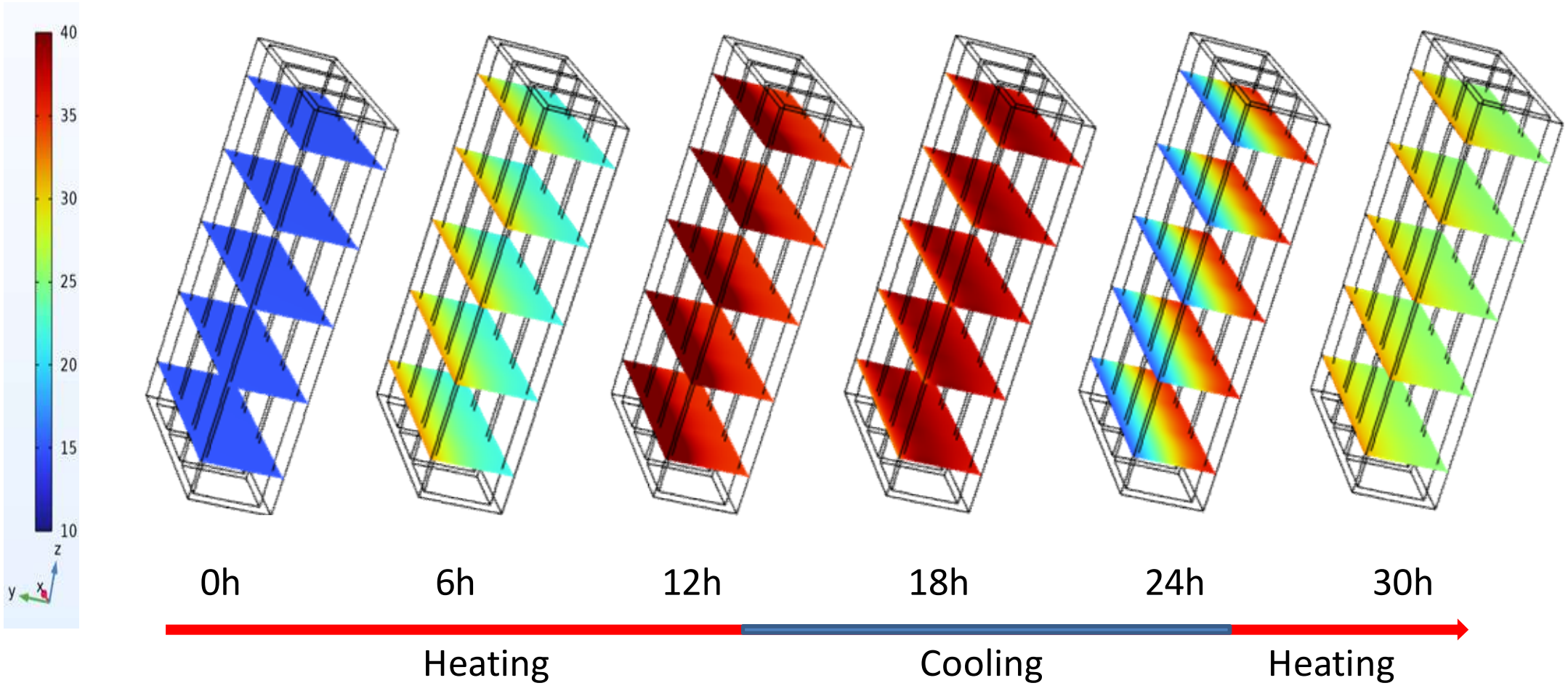

***Figure 9:*** *Temperature distribution within the PCM-filled brick at 0, 6, 12, 18, 24 and 30h, respectively.*

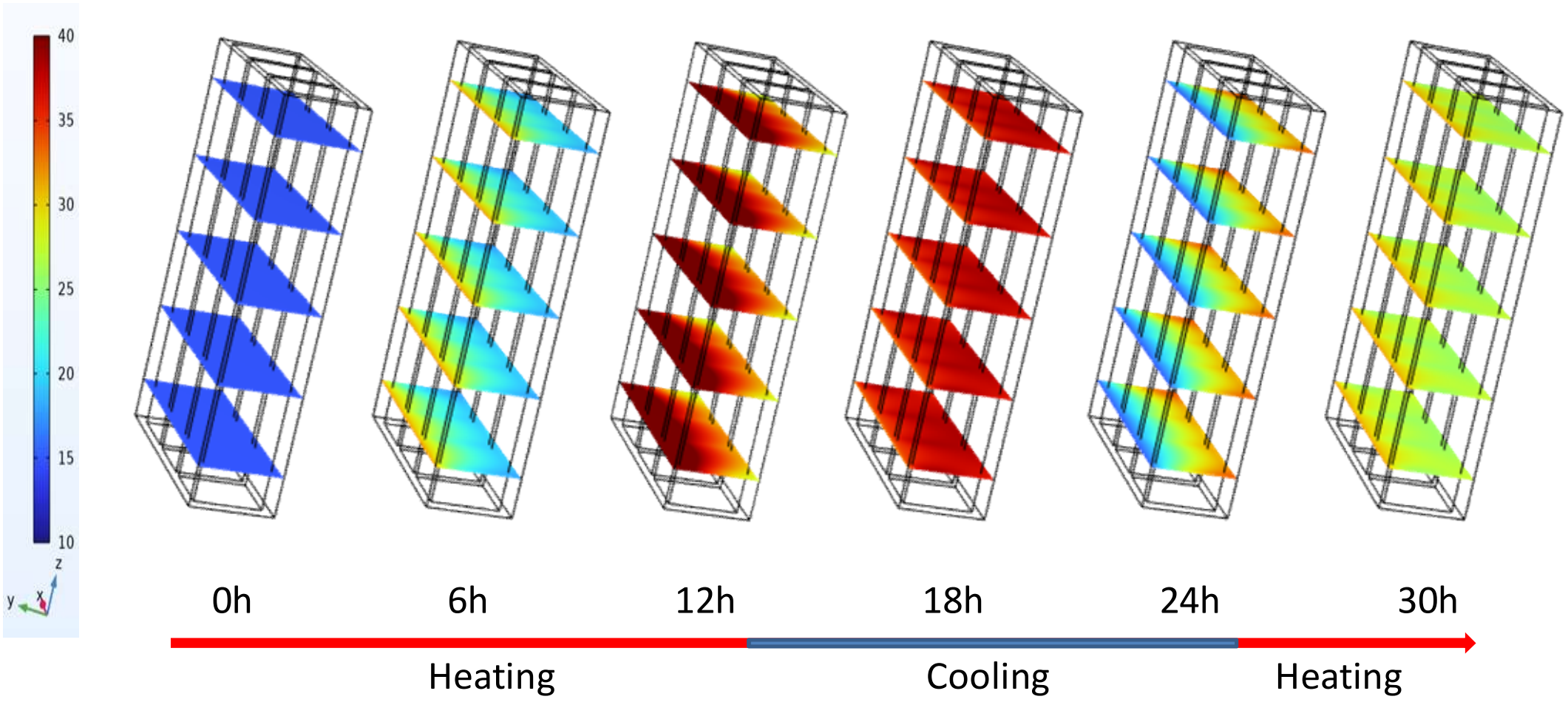

***Figure 10:*** *Temperature distribution within the brick without PCM at 0, 6, 12, 18, 24 and 30h, respectively.*

It is noteworthy that although the PCM-filled bricks exhibit good thermal inertia, the valley output temperature after the first complete heating–cooling cycle reached 24.4 °C, which is significantly higher than the crystallization phase change temperature of the PCM (22.3 °C). This indicates that the PCM did not fully complete the crystallization phase transition during the cooling phase. As a result, a portion of the PCM remains in an amorphous state at the beginning of the next heating cycle, which compromises the cooling efficiency of the phase change storage in the subsequent cycle. This issue is closely related to the quantity of the PCM incorporated within the hollow brick structure. Therefore, moderately reducing the amount of the PCM in the hollow cavities may be a promising strategy to improve the overall efficiency of the composite system.

Previous studies have demonstrated that the incorporation of appropriate additives can effectively mitigate this issue. High thermal conductivity materials such as silica and carbon black composite additives [40], oil ash [41][42], and metallic nanoparticles (e.g., aluminum, copper) [43] significantly enhance the thermal conductivity of the PCM matrix, accelerating heat transfer and facilitating a faster and more complete phase transition under dynamic thermal cycling conditions. Additionally, certain natural or waste-derived materials, such as plant fibers and waste feathers [44], although not directly improving thermal conductivity, contribute to reducing PCM content, enhancing the structural stability of composites, and lowering overall material costs due to their porous structure and low cost. These approaches help optimize the energy efficiency and economic viability of PCM-based thermal storage systems while maintaining thermal performance.

Therefore, future research should focus on the following aspects:

1) Optimal PCM loading and performance enhancement: Investigate the synergistic effects of different additives and their concentrations on the thermal properties and phase change behavior of PCMs, providing both theoretical and practical guidance for material optimization, efficiency improvement, and cost reduction.
2) Mechanical properties and lifecycle performance: In theory, since the PCM is not directly incorporated into the brick matrix but rather exists as a filler, its impact on the overall mechanical properties of the brick and the structural stability of the building is expected to be minimal or negligible. However, systematic mechanical testing and long-term durability studies of PCM composite materials, combining experimental tests and numerical simulations, are essential to provide comprehensive performance data and engineering validation to support the practical application of this technology in construction.

3) Cost control and economic benefits analysis: Although the initial investment for PCM-filled bricks is higher than that for conventional bricks, their significant energy-saving effects during the building's operational lifecycle can effectively reduce heating and cooling energy consumption, thereby shortening the payback period. From the perspective of raw materials, laboratory processes, construction techniques, and energy-saving performance, this PCM preparation technology is based on low-cost and readily available raw materials such as polyethylene glycol (PEG), hexamethylene diisocyanate (HMDI), and glycerol. At the current laboratory scale, aside from the consumption of inert gas (argon), the main energy consumption comes from the hot water circulation of the reactor envelope (60–70°C), mixer operation, and post-curing at 100°C. Compared with complex microencapsulation processes, this technology is expected to significantly reduce energy consumption and operating costs during industrial production. Furthermore, the construction stage does not require any special procedures, further lowering the overall cost. More importantly, this technology not only reduces operational energy consumption but also helps decrease greenhouse gas emissions, demonstrating considerable environmental and economic benefits. In future work, once sufficient technical and economic data from scale-up studies are obtained, a comprehensive cost-benefit analysis should be conducted to systematically evaluate the economic viability of the system.

## 4. Conclusion

This work explored a method of direct integration of a S-S PCM (PUX-1500-20) into hollow bricks used in building envelopes, aiming to enhance thermal inertia and improve indoor thermal comfort during summer, while maintaining structural integrity and PCM stability under natural drying conditions.

Based on the experimental test and numerical simulation results, the conclusions can be drawn as follows:

- The experimental results show that the PCM system required 5 hours to reach thermal equilibrium and 8 hours to cool to room temperature, significantly longer than 1 hour and 6 hours observed for the non-PCM system, indicating excellent thermal buffering performance. Meanwhile, due to the energy storage capacity of the PCM, the PCM-filled bricks effectively reduced indoor temperatures by 1–2°C compared to bricks without PCM. Overall, the use of PCM effectively enhances thermal inertia, and contributes to improved energy efficiency in buildings.
- The numerical simulation results are consistent with experimental tests: PCM-filled bricks exhibit a phase shift of 7 hours and a decrement factor of 0.38, whereas the corresponding values for bricks without PCM are only 4.5 hours and 0.39. This indicates that the incorporation of PCM can effectively reduce indoor temperatures, smooth temperature fluctuations, and delay the occurrence of peak temperatures during the heating process.
- Excessive PCM loading impedes the full realization of system efficiency. Future research should further explore the synergistic regulation mechanisms of various additives on the phase change kinetics and thermal management performance of PCMs. By integrating experimental studies with multiphysics simulation methods, alongside mechanical performance and lifecycle testing as well as comprehensive economic analysis, it is expected that material compositions and structural designs can be optimized, thereby promoting the efficient and sustainable application of PCMs in building energy conservation.

**Funding:** This work was financially supported by the National Research Agency under the France 2030 Plan, bearing the reference ANR-21-EXES-0008 within the framework of the CY Générations project.

**Credit author statement**

**Qiong YE:** Synthesis, Analysis, Investigation, Software, Data curation, Writing - original draft.

**Labouda BA:** Simulation, Data curation.

**Giao T.M. Nguyen:** Validation, Data curation, Supervision.

**Rafik ABSI:** Software, Simulation.

**Béatrice LEDESERT:** Validation, Data curation, Supervision.

**Gilberte DOSSEH:** Validation, Data curation.

**Ronan HEBERT:** Validation, Analysis, Investigation, Software, Data curation, Supervision, Funding acquisition.

All authors contributed to this manuscript.

## Declaration of competing interest

The authors declare that they have no known competing financial interests or personal relationships that could have appeared to influence the work reported in this paper.